# Enabling Hexa-X 6G Vision: An End-to-End Architecture

Bahare M. Khorsandi[1], Mohammad Asif Habibi[2], Giuseppe Avino[*,3], Sokratis Barmpounakis[4], Giacomo Bernini[5], Mårten Ericson[6], Bin Han[2], Ignacio Labrador Pavón[7], José María Jorquera Valero[8], Diego R. Lopez[9], Bjoern Richerzhagen[10], Rony Bou Rouphael[11], Merve Saimler[12], Lucas Scheuvens[13], Corina Kim Schindhelm[10], Peter Schneider[1], Tommy Svensson[14], and Stefan Wunderer[1]

[1]Nokia, Germany, [2]RPTU, Germany, [3]LIST, Luxembourg, [4]WINGS, Greece, [5]Nextworks, Italy, [6]Ericsson, Sweden, [7]Atos, Spain, [8]University of Murcia, Spain, [9]Telefonica, Spain, [10]Siemens, Germany, [11]Orange, France, [12]Ericsson, Turkey, [13]TU Dresden, Germany, [14]Chalmers, Sweden

*Abstract* —The end-to-end (E2E) architecture for the 6th generation of mobile network (6G) necessitates a comprehensive design, considering emerging use cases (UCs), requirements, and key value Indicators (KVIs). These UCs collectively share stringent requirements of extreme connectivity, inclusivity, and flexibility imposed on the architecture and its enablers. Furthermore, the trustworthiness and security of the 6G architecture must be enhanced compared to previous generations, owning to the expected increase in security threats and more complex UCs that may expose new security vulnerabilities. Additionally, sustainability emerges as a critical design consideration in the 6G architecture. In light of these new set of values and requirements for 6G, this paper aims to describe an architecture proposed within the Hexa-X, the European 6G flagship project, capable of enabling the above-mentioned 6G vision for the 2030s and beyond.

*Keywords—E2E architecture, sustainability, security, 6G*

## I. INTRODUCTION

As the world embraces the full potential of the fifth generation of mobile network (5G), a new era of wireless communication is already on the horizon. 6G architecture promises to elevate connectivity to unprecedented heights and empower a multitude of emerging technologies that will redefine the way we communicate [1]. At its core, the architecture of 6G is envisioned to be an intricate network of networks, intertwining satellites, terrestrial infrastructure, and beyond. It will transcend the traditional boundaries of connectivity, encompassing a wide range of frequencies, including the sub-terahertz and terahertz bands, to facilitate data transfer at much higher speeds. It is expected that the architecture of 6G will harness innovative techniques to reduce power consumption and carbon footprint. As sustainability becomes a global imperative, 6G technology will strive to balance the quest for speed and connectivity with environmental, societal, and economical responsibility.

The design of 6G architecture also requires accommodating the requirements of the new UCs with high availability of resources, reliability, and extreme precision. A subset of UCs from Hexa-X [1], focusing on *dependability* has been selected in this paper to showcase the technical design reasoning for the proposed architecture. Dependability attributes characterizing the application demands for the respective service. A brief description of these UCs can be found in the following and Fig. 1.

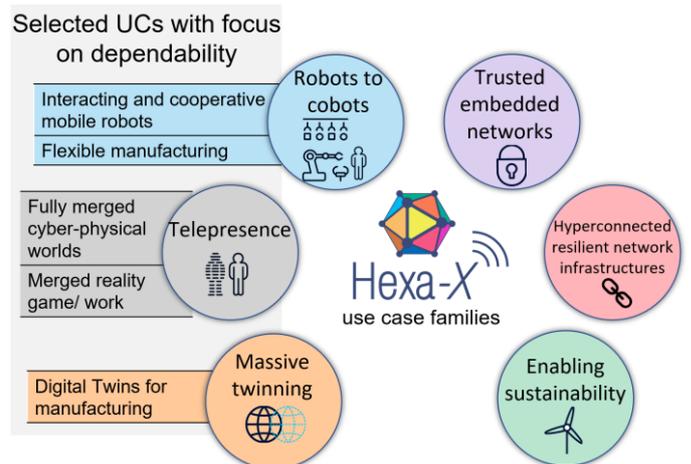

*Fig. 1: Hexa-X dependability sub-set UCs*

The "*Robots to cobots*" family prioritizes network flexibility in the face of mobility, while also addressing stringent availability and reliability requirements, coupled with low latencies for direct communication among cooperating mobile robots. Within the "*Telepresence"* family of UCs, the focus lies on the inclusion of novel Human Machine Interfaces (HMIs) and the associated demands in terms of bandwidth and low latency. For the "*Massive Twinning*" family, the potential of a digital representation of the manufacturing environment combined with a more detailed digital twin of the network environment is being studied. This enables the assessment of the impact of human presence on the next generation of communication system, an important building block for human-machine collaboration in flexible manufacturing environments [2].

The rest of this paper is organized as follows: We commence by discussing the requirements of several industrial use cases for 6G in Section II. Then, we highlight key design values that need to be considered as a design principle for 6G architecture in Section III. Following that, we propose an E2E architecture for 6G in Section IV. Finally, we conclude the achievements of this paper in Section V.

---

## II. REQUIREMENTS OF INDUSTRIAL 6G USE CASES

This section focuses on the main requirements for Hexa-X vision which motivate the technical enablers and the design of the E2E architecture.

**Inclusion** refers to the capability of using and the ease of access to a service. This encompasses both technical aspects, such as coverage (i.e., spatial, and temporal availability of a network and service), and user-centric aspects, such as availability and accessibility of the service for a specific group of people, e.g., through novel HMIs or interaction with robots. Consequently, the term "inclusion" is used interchangeably with "digital inclusion." Inclusion is a key aspect of the "*fully merged cyber-physical worlds*" and the "*merged reality game/work*" UCs, given that they enable participation and interaction in the virtual (or augmented) world, even if physical barriers (distance, safety of a workplace, disabilities) would hinder those in the real world.

**Flexibility** refers to the ability of the system to adapt to (or undergo adaptation to) changes in its environment and utilization, considering the costs that such changes would inflict. This attribute is closely intertwined with the dependability and resilience of a system. Flexibility also impacts the sustainability of the system, as it can allow the system to address additional (and potentially still unknown) UCs in a resource- and cost-efficient way through re-use and re-configuration of components. Flexibility also refers to being able to integrate different types of networks (e.g., mesh or Ad-hoc networks) into the overall network topology. Flexibility is a key requirement in the "*interacting and cooperative mobile robots*" and "*flexible manufacturing*" UCs, as the system must adapt to mobility and reconfigurations of the manufacturing environment.

**Trustworthiness** spans from security to aspects that are covered by the dependability framework, such as availability and reliability of a communication service. The overall range of factors contributes to creating a trustworthiness level that enables a larger set of UCs that wouldn't be possible with lower privacy or lower network availability levels.

**Sustainability** considers two aspects: sustainability of the 6G system itself (e.g., environmental footprint, emissions, and energy consumption) and the enablement effect 6G systems have on different sectors.

More in-depth details on trustworthiness and sustainability aspects are given in Section III.

## III. KEY DESIGN VALUES

### A. Security, privacy and trustworthiness

The 6G network confronts a comprehensive threat landscape. Beyond persistent challenges, such as fragile Information and Communication Technology (ICT) infrastructure, susceptible software and cloud configurations, and operational deficiencies, there is the continually evolving cyber-attack paradigm to consider. Distinctly, 6G brings its specific risks, stemming from upcoming trends and technologies unique to this generation. These concerns are particularly acute for dependability-critical applications. Due to their rigorous reliability criteria, such applications are especially vulnerable to any form of network disruptions, no matter how trivial. Often entailing critical processes and safeguarding sensitive data, a cybersecurity breach within these realms could precipitate substantial losses, jeopardizing life, safety, privacy, and material assets. Their increasing convergence with human-cyber-physical systems augments complexity, necessitating a heightened reliance on data and Artificial Intelligence (AI)-driven methodologies. This convergence expands the potential attack surface and ushers in fresh vulnerabilities. A more detailed exposition of these security concerns can be found in [1].

Trustworthiness is intricately linked to both security and privacy. While these factors are pivotal, they often remain opaque to the service user. Building trust necessitates a protracted period of consistent, undisturbed service devoid of any security and privacy infringements. To proactively create trust, black-box solutions must be avoided. Approaches to achieving an initial foundation of trust involve presenting security certifications by trusted entities for system equipment, embedding security protocols throughout the supply chain, elucidating insider attack mitigation strategies, and ensuring adherence to secure operational practices. Defining precise Key Performance Indicators (KPIs) for security, privacy, and overall trustworthiness remains a complex endeavour. Preliminary metrics to consider in this context include:

- Strength and quantum-resilience of cryptographic algorithms.
- Rigor of security policies, exemplified by patterns like re-authentication.
- Coverage of security specifications concerning interfaces and services.
- Extent of security assurance coverage.
- System's robustness against known attacks targeting standardized protocols and procedures.

Obviously, such KPIs are much vaguer than classical KPIs but still, we consider them useful to get an impression of the security posture of future 6G standards, technologies, and architectures.

As a concrete approach to address the trustworthiness KVI, we propose to utilize a Level of Trust Assessment Function (LoTAF) that plays a key role in enhancing security by analyzing the use of applicable security technologies. It serves as a neutral and bidirectional service, catering to both trustors (users), those whom it aids in making informed decisions, and trustees (network providers), those whom it offers insights into compliance with security requirements and opportunities for improving service quality.

To ensure effectiveness, the LoTAF identifies relevant technologies, analyzes risks, and investigates potential attack patterns. It also leverages AI-driven threat detection mechanisms and human experience to enhance the identification of cybersecurity threats and attacks. Besides, Machine Learning (ML)-based intelligent optimization algorithms enable a balanced configuration of network services, considering factors like cost and risk based on user-defined criteria.

The LoTAF categorizes the trustworthiness of network services using a set of thresholds, assigning Level of Trust (LoT) values to each category. This allows users and providers to make informed decisions and determine the affinity between users' security and privacy intents and providers' capabilities. Therefore, LoTAF is designed as a service that belongs neither to network providers nor users, but as a neutral service that could be deployed, e.g., in a decentralized marketplace where on-demand resources and services may be provided. In addition, a continuous update mechanism ensures the adaptability of the LoT assessments, incorporating feedback from users and reassessing based on detected attacks. Overall, the LoTAF constitutes a new and agile way to enhance the security and privacy of 6G network services, empowering users and providers with valuable insights and decision-making support.

*B. Sustainability*

The E2E architecture serves as a foundational element in addressing the sustainability challenges across economical, societal, and environmental pillars. In the following we presents the result of our work to address such challenges:

**Economical sustainability**: The baseline that we considered in the analysis of Total Cost of Ownership (TCO) is 5G New Radio Standalone (5G NR SA). A study conducted by the Global System for Mobile Communications Association (GSMA) in [3] compares the TCO between 4G and 5G considering five key cost items: (i) Radio Access Network (RAN) infrastructure, (ii) energy consumption, (iii) Core Network (CN) infrastructure, (iv) backhaul, and (v) other network-related expenses (e.g., personnel, network management).

To compare costs between 5G NR SA and 6G, five key technical enablers have been identified. These include intelligent networks (User Equipment (UE), programmability, dynamic functions, AI as a Service (AIaaS)), flexible network integration (ad-hoc networks, Device to Device (D2D), mesh), efficient networks (RAN/CN signaling, function refactoring, compute as a service), 6G RAN features (high data rate links, Distributed-Multiple Input Multiple Output (D-MIMO), localization), and service management features focusing on continuum management, orchestration, and AI-driven orchestration. The impact of each enabler on costs depends on the specific UC. We focus on the "*Fully Merged Cyber-Physical Worlds*" UC. The analysis quantitatively assesses cost reductions, focusing on 6G RAN and service management. Our initial TCO qualitative analysis indicates that 6G RAN enablers exclusively impact "RAN infrastructure" costs, while service management enablers affect "energy consumption" and "other network-related costs." While 6G RAN enablers have the potential to halve the costs of the RAN infrastructure [4], allocating the 10% reduction from service management enablers [5] requires consideration of their baseline weights (39% for "energy consumption" and 7% for "other network-related costs" [3]), resulting in a 1:5.6 ratio. Consequently, service management enablers reduce "energy consumption" by 8.5% and "other network-related costs" by 1.5%. Fig. 2 compares 5G NR SA [3] and 6G TCO with two enabler families, illustrating a notable 26.4% reduction.

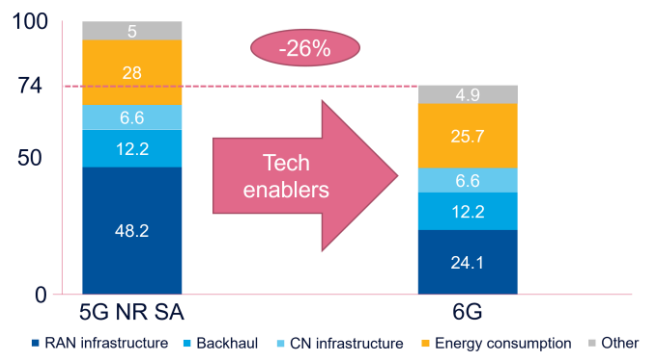

*Fig. 2: TCO comparison between 5G NR SA and 6G for the "Fully merged cyber-physical worlds" UC.*

**Societal sustainability:** The ICT sector's footprint impacts include direct effects from infrastructure and devices across their lifecycle, and indirect effects from ICT-enabled solution deployment. The enablement effect relates to Greenhouse Gas (GHG) emission reductions from implementing an ICT solution [4]. Assessing 6G's enablement effect involves comparing it to a baseline emission scenario. [6] provides a standard method to quantify the enablement effect of existing technologies but lacks guidance for assessing future ones. This paper offers comprehensive guidelines for over 30% $CO_2$ emission reductions in 6G sectors. Various sectors can benefit from solutions like remote industrial control and professional consultation. Implementation requires strategic policies for cultural shifts and enhanced collaboration tools. Effective training and fiscal support are essential for maximizing adoption. Assessing Mixed Reality (MR) UCs and its adoption, challenges persist. Motion sickness, limited mobility with MR headsets, and low-resolution visuals are common issues. These problems, often called "cyber-sickness", result from delays between real-world and virtual-world movements [7]. 6G's low latency and edge computing reduce headset weight and enable lighter batteries for wireless MR, while also enhancing MR security with robust protocols and encryption for sensitive applications.

**Environmental sustainability:** While 6G enhances classic KPIs like peak throughput, it strives to reduce energy consumption per transported data unit (Wh/bit), or Energy Efficiency (EE), by 90%, in line with past trends. The goal is "quasi-zero watt at zero load", achieved through dynamic scaling of RAN infrastructure based on real-time traffic load. This energy flexibility relies on highly reconfigurable sub-systems (hardware/software) adaptable to various scenarios. Assessing EE involves DC power and data volume measurements over a specified duration, with ETSI providing a standardized methodology [8]. However, early 6G design phase EE assessment is complex, requiring model-based evaluations of components. Defining the baseline, such as 5G NR SA with high (80%) and low (20%) traffic loads, is crucial. For domain-specific comparisons, additional details must be considered, like a 64 TRx setup on FR1 with a 100 MHz bandwidth to analyze a 6G antenna array and RF front-end performance. Key technology drivers for 6G's tenfold EE improvement include but not limited to: (i) RF power amplifiers (ii) Electronic components (iii) AI, multi-goals optimization, and adaptive air interfaces (iv) Sleep modes (v) Path loss (vi) Infrastructure sharing. While these levers focus on the RAN's impact, other parts of the 6G system should also contribute to the tenfold EE reduction.

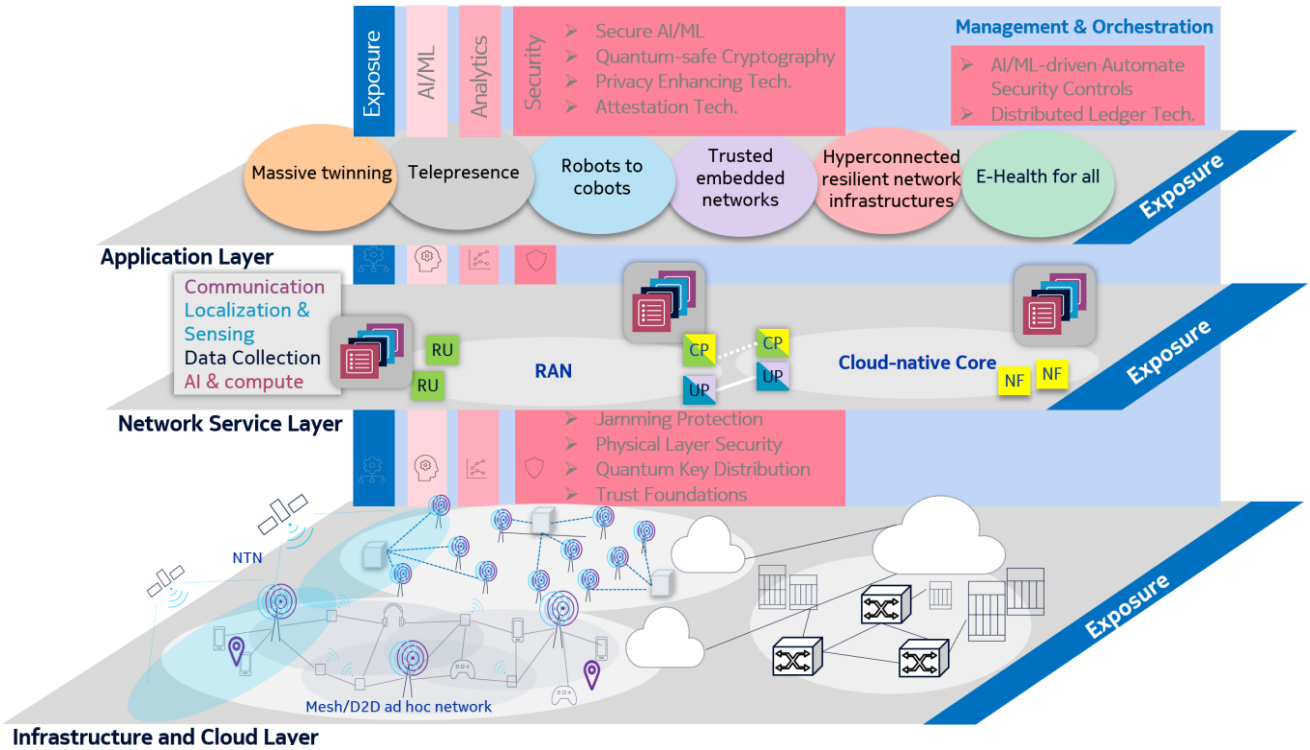

Fig. 3: Hexa-X E2E architecture

IV. END-TO-END ARCHITECTURE

Fig. 3 presents an update on the Hexa-X proposed E2E architecture as well as the technical enablers in a layered structure composed of *Infrastructure and cloud layer*, *Network service layer,* and *Application layer* as well as a cross-layer *Management and Orchestration (M&O)* functionality based on [9]. Fig. 3 also demonstrates AI/ML, analytics and security as features which effecting all layers and elements in the architecture. Following subsections presents each of layers and relative architectural enablers in more details.

*A. Infrastrructure and cloud layer*

This layer plays a crucial role in enabling the seamless communication experience that 6G promises to deliver. It provides physical and virtual resources to host network services, cloud applications and application layer [10, 11]. This layer is made of a set of technical enablers that work together to support the ultra-high speed, reliable, and secure communication network. At its core, it comprises a network of interconnected devices, including Internet of Things (IoT) and UE, Base Stations (BSs), small and macro cells, Access Points (APs), cloud infrastructure, etc. The *Infrastructure and cloud layer* also includes the integration of extreme-edge, that is part of a network with high heterogeneity of devices, characterized by a wide variety of technologies, in terms of both hardware and software. Having extreme-edge as part of the E2E network cloudification can lead to the full integration of cloud-native technologies, such as distributed computing and virtualization. The selected key 6G technologies for this layer for further elaboration are the *D-MIMO, Localization & Sensing and NTN*.

**D-MIMO** enables the use of multiple antennas at both the transmitter and receiver, leading to improved data transfer rates, increased spectral efficiency, coverage extension, and enhanced communication **reliability**. Further on, D-MIMO has the potential to allow for further densification of Radio Units (RUs) for increased and consistent area capacity; mitigate shadowing and blockage thanks to macro diversity; achieve sufficient link margin despite output power limitations and high pathloss at high frequencies; and allow for lowering Effective Isotropic Radiated Power (EIRP) that can simplify deployment. While D-MIMO is not yet supported within 3GPP 5G standards, Multi-Transmission Point (multi-TRP) offers a foundation. Real-world D-MIMO deployment faces challenges in architecture, Central Processing Units-AP (CPU-AP) split, fronthaul/backhaul, scalability, and precoding. These issues necessitate scalable digital-analog approaches and efficient, integrated backhaul/fronthaul solutions. Given the importance of densification at high frequencies and available spectrum, cost-effective, decentralized solutions are prioritized over spectral efficiency initially. At low frequencies, the demand for spectral efficiency favors more centralized, digital strategies. E2E architecture should accommodate diverse D-MIMO nodes, with a clear separation of control and user planes, enhancing sustainability by aligning with service demands. Multi-connectivity and CPU-AP splits also raise security concerns, particularly in authentication and MAC layer splitting, suggesting a move towards centralizing user plane security to enhance network scalability and reduce vulnerabilities.

*Localisation and Sensing* is one of the new services envisioned for 6G, residing in the infrastructure layer but also supported by network functionality in the network service layer. One major differentiator of 6G compared to previous generations of mobile communication is the vision of integrated sensing and communication. Based on sensing information, location- and context / environmental-aware services can emerge to form an ecosystem of services and applications which may even include network external sensors and localisation systems. The challenge from an E2E architecture perspective is the integration of all technical enablers both regarding communication and sensing, and the management of each hardware/software component to enable data and information flow. Well defined interfaces on hardware, protocol and semantics level allow for new services and applications in the field of location and environmental context to emerge and an ecosystem of hardware vendors, service, and application developers and providers to flourish.

*Non-Terrestrial Network (NTN)* is to provide global coverage and enabling communication in remote and underserved areas. As mentioned earlier, an important target of future 6G networks is the **inclusion**. This goal necessitates a flexible NTN architecture for efficient and sustainable 6G. NTNs can seamlessly integrate with Terrestrial Networks (NTs), employing transparent architecture where 5G BS at Ground Stations (GS) relay data via satellites directly to UE. The role of Inter-Satellite Links (ISLs) is pivotal for extending satellite coverage, requiring an architecture that supports efficient signal processing for comprehensive global coverage, including oceans. This involves managing ISLs effectively, especially given the dynamic nature of satellite movements, which necessitates frequent setup and teardown of links, posing challenges in 5G/6G networks. To enhance network flexibility and counteract the inherent rigidity of NTNs, mesh ad hoc network topologies are crucial for the evolution towards 6G.

### B. Network service layer

This layer is responsible for providing various services to end-users [10]. The layer also depicts Network Functions (NFs) and their services that are used within the network and are not exposed to the end-users. This layer will ensure that users have access to high-quality, reliable, and secure services [11]. In addition to the communication services, new services such as AI and compute, analytics, data collection and localisation and sensing are raising with 6G. NF, operations, and services can be implemented as cloud-native microservices, leading to a more softwarised, intelligent, and efficient 6G architecture.

The selected key 6G technologies for this layer for further elaboration are *AIaaS* and *Ad-hoc networks*.

In 6G, **AI** can enhance in-network enablers such as M&O, physical layer and resource management, while AIaaS delivers on-demand AI-related network services, democratizing AI access [11]. 6G offers AI tools as a service, augmenting the overall network performance (i.e., improving the **inclusion**) and may improve other key values such as **sustainability**. The necessary AI functionality is included in the *network service layer*, except the exposure part of AIaaS which belongs to the *application layer*. As shown in Fig. 4 an AIaaS framework can be considered and integrated into the 6G architecture to expose and offer AI capabilities to a wide range of consumers. These consumers include M&O, other network functions (belonging to different domains, such as RAN and core), application functions, as well as third parties [12]. The objective is to provide a suite of AI functions, including training and monitoring, with tailored inference capabilities that depend on the specific consumer's needs. This supports the implementation of closed-loop network and service automation.

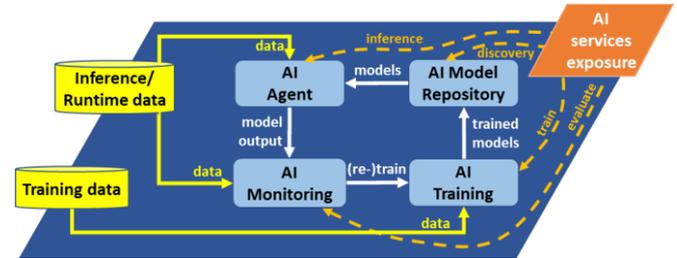

*Fig. 4: AIaaS framework*

The proposed AIaaS framework comprises 4 functions: *AI model repository* to catalog trained AI/ML models for deployment. *AI training* function is dedicated to training ML models and generating executable versions for integration into AI agents. *AI monitoring* function assesses the performance of ML models, triggering retraining when necessary. *AI agent* employs trained ML models for inference and data pre-processing. Furthermore, the framework incorporates two centralized data stores. One to collect and expose inference data, facilitating the runtime operation of ML models. The second stores training data essential for AI training purposes. This separation allows for versatility in accommodating diverse data sources and processing needs.

The **Ad-hoc network** is created and controlled by a management network. The primary functionalities that are addressed comprise selecting nodes and extreme-edge devices for **inclusion** in the ad hoc network formation, assessing the trust level of a node for participation in the D2D/mesh network, and unifying the modeling of extreme-edge nodes and devices in terms of their network and computational resources, capabilities, and constraints. It also includes integrating these elements with network and service orchestration for seamless management, control, and enforcement, as well as discovering nodes and extreme-edge devices, which encompasses synchronization aspects for advertising capabilities.

Fig. 5 illustrates the gains in terms of communication resources utilization ratio, energy consumption, and cost in a public safety scenario. A disaster took placed, and timely allocation of excessive communication and computing resource in a remote area is required. A number of teleoperated robotic platforms need to navigate and provide real-time video service for further processing. The scenario configuration involves a set of traffic sources, located at various locations in a remote area (e.g., rural area), a set of candidate flexible nodes with access and backhauling capabilities (Unmanned Aerial Vehicle (UAVs) are assumed in this scenario), as well as computing capabilities. There are predefined cost vectors associated to each configuration and network node. The results compare the KPIs in the case of a static network infrastructure, i.e., a macro cell-based deployment (blue plot), against the flexible allocation of a set of UAV-powered network formation (green plot). According to the results, certain trade-offs need to be considered, depending on the amount of resources that need to be allocated; for example, for very high number of UAVs the cost may exceed the static infrastructure costs; nevertheless, there are clear gains in terms of energy consumption when commissioning only the required number of flexible network nodes.

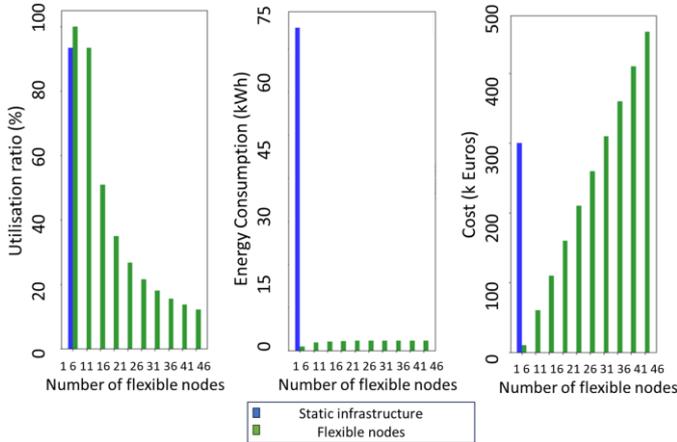

Fig. 5: Comparison of static infrastructure with flexible UAV-based nodes in a disaster scenario.

## C. Application layer

The topmost layer is *Application layer* which interacts directly with end-user applications, facilitating the exchange of data and information [12].

## D. Management and Orchestration (M&O)

The M&O functionality, depicted as a vertical blue rectangle in the architecture, spans three layers to orchestrate resources at the infrastructure level and manage services at the network service layer for application layer services. It aims to enhance **flexibility** and **trust** through a "device-edge-cloud" continuum, extending 5G network slicing to include all network resources and even those beyond the MNO's (Mobile Network Operator) scope [10]. This approach integrates extreme-edge and addresses the volatility of device connections. To support service and resource orchestration, M&O combines core capabilities with AI/ML for predictive actions, security across levels, and advanced monitoring for comprehensive analytics, thereby facilitating orchestration decisions.

## E. Security Architecture

The 6G E2E architecture must be accompanied by a security architecture specifying what security and privacy mechanisms must be applied, and how they interact to ensure overall security, privacy and **trustworthiness**. Essential 6G security technology enablers and their placement in the architecture is highlighted in Fig. 3. Secure AI/ML, quantum-safe cryptography, and privacy enhancing technologies apply to functions and procedures on all layers. Security and risk management must evolve towards AI/ML-driven automated closed loop mechanisms. Distributed ledger technologies may facilitate distribution of trust in interdomain management. To ensure system integrity at boot and runtime, attestation technologies must be applied, building on trust foundations, such as trusted execution environments, in the secure infrastructure and cloud layer. Further in this layer, mechanisms must be applied to mitigate the inherent threat of jamming attacks. Moreover, physical layer security technologies and quantum key distribution may complement the cryptographic schemes used today.

## V. CONCLUSION

This paper provided the Hexa-X vision on 6G E2E architecture. It developed based on 6G core values namely sustainability, inclusion, and trustworthiness in order to satisfy the UC requirements. The design of the 6G architecture is flexible and highly specialized to accommodate a comprehensive and highly advanced wireless ecosystem. The main enablers for improved flexibility and inclusions described in this paper are D-MIMO, NTN, and ad-hoc networks. The sensing and AIaaS enablers extend the mobile network for new 6G UCs and, finally, the M&O enabler using the cloud continuum improves flexibility and trustworthiness.

ACKNOWLEDGEMENTS

This work was supported by the European Commission through the H2020 Project Hexa-X under Grant 101015956.